# Selectively embedding multiple spatially steered fibers in polymer composite parts made using vat photopolymerization

Vivek Khatua[1], B. Gurumoorthy[1,2], G. K. Ananthasuresh[1,2]

[1] Centre for Product Design and Manufacturing, [2]Department of Mechanical Engineering,

Indian Institute of Science, Bengaluru, Karnataka, India

{vivekkhatua, bgm, suresh}@iisc.ac.in

**Abstract**

Fiber-Reinforced Polymer Composite (FRPC) parts are mostly made as laminates, shells, or surfaces wound with 2D fiber patterns even after the emergence of additive manufacturing. Making FRPC parts with embedded continuous fibers *in 3D* is not reported previously even though topology optimization shows that such designs are optimal. Earlier attempts in 3D fiber reinforcement have demonstrated additively manufactured parts with channels into which fibers are inserted. In this paper, we present 3D printing techniques along with a printer developed for printing parts with continuous fibers that are spatially embedded *inside* the matrix using a variant of vat photopolymerization. Multiple continuous fibers are gradually steered as the part is built layer upon layer instead of placing them inside channels made in the part. We show examples of spatial fiber patterns and geometries built using the 3D printing techniques developed in this work. We also test the parts for strength and illustrate the importance of spatially embedding fibers in specific patterns.

## 1. Introduction

Fiber-reinforced composites have mechanical properties superior to the individual properties of fiber and matrix without having to chemically alter either of them [1]. Since we are interested in stiffness and strength against failure, we consider fibers that have high modulus and strength in its longitudinal direction. The matrix in composites can be metal, polymer, or ceramic. Fiber-reinforced polymer composites (FRPCs) are lightweight and stiff and hence are preferred in many engineering applications. Although addition of fibers in random or regular patterns has been shown to improve stiffness, continuous fibers steered and embedded in the matrix show much higher stiffness [2] and increased failure strength [3] for given weight and cost. In [2] and [3], fibers were oriented and placed along the principal stress directions to behave as the primary load-bearing members of the part while the matrix helps in retaining the orientation of the fibers. Since the fibers need to be the load-bearing members for stiffness, they ought to be oriented and placed in specific patterns in a given part. This is because orthotropic behavior of fibers is beneficial when effectively utilized.

Many composite manufacturing techniques have freedom and flexibility of orienting and placing multiple fiber tracks. However, they are suitable for specific part geometries (e.g., volumes of revolution that can be filament-wounded or shells with embedded fiber mats) and are not amenable for freeform part geometries. 3D printing, an additive manufacturing method, is a recent development to embed continuous fibers in desired orientations and locations inside the matrix along with freeform fabrication of geometry. All such inclusions of fibers are restricted to in-plane deposition or on a curved surface. Selectively embedding continuous fibers in 3D is largely unexplored.

We demonstrate a 3D printing technique to embed continuous fibers that continue along the build direction of a layer-by-layer process of matrix polymerization using stereolithography. Before describing the new technique, we review the related work.

We first discuss textile sewing method for fiber mats and filament-winding techniques. Takezawa et al. [4] used a stitching method called Tailored Fiber Placement (TFP) to fabricate fiber-mats with oriented continuous carbon fibers for Vacuum-assisted Resin Transfer Molding (VaRTM). The orientation of fiber from their work demonstrated stiff, winkle-free doubly curved surfaces. Others have fabricated stiff structures using TFP along with their customized resin transfer processes [5, 6] and have presented potential applications of sewing processes for fabricating steered fiber composites [7-10].

Robotic Filament Winding (RFW) and Coreless Filament Winding (CFW) are extensions of traditional filament winding processes possessing enormous process flexibility in laying continuous fibers. Robotic filament winding (RFW) has added degrees of freedom at the layup head compared to conventional filament winding. Sorrentino et al. [11] used RFW to make a lattice structure encapsulated in a cylindrical shell (iso-grid structures) which improved the mechanical performance of a fiber-reinforced cylinder. They also showed a practical application of a part of complex geometry and fiber path made using RFW [12]. Although this process is dependent on a mold, they ensure compaction of fiber-matrix and thorough impregnation of fibers with matrix. Whereas Coreless Filament Winding (CFW) liberates the process from mandrel and molds by using a conformable head and tail stocks, the path planning is complex. CFW can make complex structures with multiple fiber materials providing material heterogeneity and programmable anisotropy [13,14]. Since all the layup and windings are on exposed surfaces, embedding fibers inside the matrix (for example, a helical fiber inside a solid cylinder) is not possible.

On the other hand, recently 3D printing techniques can produce freeform geometries that include continuous fibers. Few of the existing fiber-reinforced 3D printing techniques that have polymer matrix are elucidated next to delineate strengths and limitations in placing and orienting fibers.

Fused Filament Fabrication (FFF) is an extrusion-based 3D printing technique where the source material is continuous thermoplastic filament. The convenience in storage and handling of the filament and embedding fibers makes FFF a preferred choice for 3D printing FRPCs today. In FFF for FRPCs, an extrusion nozzle that deposits polymer is modified to include a continuous fiber in either of the formats: unimpregnated fibers or polymer-impregnated fibers. Impregnated fibers with good interface strength and reduced inter-fiber voids have been reported compared to unimpregnated fibers [15]. This has attracted many researchers to present multiple optimized fiber placement strategies for stiff parts. All reported works on fiber reinforcement in FFF are with fibers arranged in a plane, planar [15-22] or curved [23]. In these techniques, there is a possibility of a few fibers delaminating from the part rather than breakage of the part. This indicates the failure of polymer-polymer interface of impregnated fiber coextruded with polymer, followed by polymer-fiber interface. This problem is more pronounced when the polymer filament is different from the polymer used to impregnate the fibers. In such a case, high stiffness and strength of the fibers is not fully exploited. This limitation is overcome in our technique, as described later in the paper.

Since we use vat polymerization, we present relevant prior work. The device that uses photocurable resins in a vat to 3D print a part is known as a Stereolithography Apparatus (SLA) [24]. Continuous fiber inclusions in SLA are seen with woven fiber-mat introduced in a bottom-up SLA [25, 26]. In a bottom-up SLA, the build platform starts from the bottom of the vat where a transparent screen projects a pattern and selectively solidifies a layer of the part through curing [27]. The part is made layer upon layer as the build plate moves up. Since build platform can apply pressure on to the transparent screen, woven fiber mats can be immersed, and impregnated with resin thoroughly [28]. This improves fiber-matrix adhesion as compared to FFF. Manual addition of a sheet of fiber-mat distances the process from near-net fabrication which in turn requires postprocessing for intricate geometries. Continuous fibers as an oriented fiber-track in SLA was also explored [28–31]. With selective reinforcement in matrix, Renault, and Ogale [28, 29] and Ogale and others [30, 31] reported threefold increase in yield-strength using carbon fiber at a 30% volume fraction. In this case, there is only fiber-polymer interface as compared to FFF where polymer-polymer interface of fiber filament to matrix is common. Although the results were promising, fewer advances have been made since 2001. Recently, Lu et al. [32] presented a method in vat polymerization to introduce continuous carbon fibers. Their setup uses a top-down SLA system from Gizmo Inc., and they make a cavity for placing polymer impregnated continuous carbon fiber filaments. These filaments are easy to handle as they do not splay in the presence of the resin. However, they are restricted to bending up to a limit radius of curvature. Hence complex and tortuous paths with high curvature are difficult to manufacture without breaking few fibers in the fiber bundle. This may restrict the method to simple fiber paths continuing in the build direction.

In all the work discussed so far, importance is given to the orientation of continuous fibers in the 3D printed structures and not their placement. Orientation and placement of fibers becomes relevant when we consider computational design techniques, topology optimization in particular [33, 34] that aim for lightweight and enhanced performance. Optimized designs of [33, 34] (Figure 1 a., b., and c.) have intricate spatial patterns of fibers embedded *inside* the matrix. Such structures evidently are difficult to manufacture with all the manufacturing methods presented in this section. On the other hand, optimized designs reported in [35-39] have fibers everywhere in the part. So, they can be 3D printed with the existing techniques. Our focus is on being able to 3D print the 3D designs reported in [34]. A representative 3D part we propose in this work is pictorially depicted in Figure 1c.

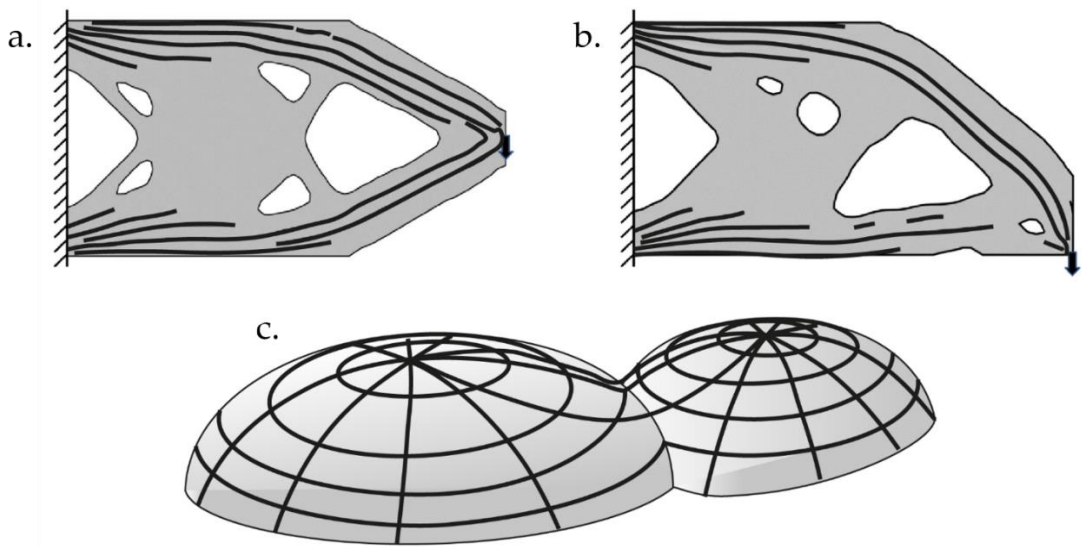

*Figure 1 a. and b. show 2D selective fiber placement as cited in [33] and c. shows a half peanut shell, which is proposed as a representative part, with the black lines showing the direction of fibers placed in principal stress direction for compressive loads.*

Brooks and Moloney [40] were one of the first to attempt spatial positioning and orientation of continuous fibers in 3D printing but they made a provision with channels while printing the part, and later inserting continuous fibers manually in the channels. They advocate subsequently dipping the part in uncured resin and curing it after the resin flows into the channels that now have fibers inserted. There is a challenge here in ensuring the flow of the resin into all channels and then sufficiently curing it. Inserting flexible and rigid fibers into tortuous channels is another challenge.

Incidentally, the case for multiple spatially steered fibers can be made by considering some of the natural structures that have fiber pattens judiciously placed in specific regions with functional fiber orientations. For example, the fibrous venations in a leaf are oriented and placed only in few places that contributes significantly to provide enhanced stiffness (to stay flat) and strength (to stay connected) compared to the lamina of the leaf [41]. A leaf would deform by its self-weight without its mid-ribs and venations. Such optimized fiber arrangements are also seen in shells of nuts and seeds, of which peanut (called groundnut in India) is an example wherein the fibers provide strength under compression. Examples from nature given here have fibers on flat or curved surfaces with heterogeneous fiber patterns. The emerging computational optimized designs not only have heterogeneous fiber patterns but also specific spatial steering in freeform 3D parts.

Thus, in this work, we aim to demonstrate realization of multiple spatially steered continuous fibers embedded inside the polymer matrix in 3D printing. We developed a custom top-down vat-polymerization system and manually embedded fibers by gradually steering them spatially as the part is made layer upon layer. We tested the parts and analyzed their behavior, as explained in Section 3.

## 2. Materials and Methods

In this section, we describe the materials used in the 3D printing process and printing processes for building different parts. The parts with fiber orientation chosen for 3D printing can be printed in multiple ways, which we discuss. Such processes cannot be realized with conventional layer-by-layer approach of placing fibers. Intricate fiber patterns need unconventional approach in 3D printing.

### 2.1. Materials

The epoxy resin used is an acrylate-based epoxy with photo-initiator sensitive to 365 nm to 405 nm wavelength of light (ANYCUBIC$^{TM}$ Clear) [42]. The solidified epoxy part has an elastic modulus of 1.5 GPa and a tensile strength of 36-45 MPa. The carbon-fiber tow used are provided by Marktech Pvt. Ltd., manufactured by Toray. The carbon fibers have a tensile modulus of 230 GPa and a tensile strength of 3.53 GPa. All the mechanical properties presented are claimed as per the manufacturers.

### 2.2. Characterization apparatus

We used Instron-5697 system with a load cell of 2 kN for testing the stiffness and strength of the 3D printed samples. An Olympus IX81 optical transmittance microscope was used for microscopic images of fiber and matrix interface. This microscope has a 40X magnification capability.

### 2.3. Custom-built 3D printer with robot in tandem to embed fibers

The SLA system has a projector procured from Optoma (Optoma x400) that uses digital mirror devices to project an image. This commercial projector is modified to emit a spectrum of 365 nm to 900 nm. Each layer is realized as a binary bitmap projection (black and white) of which only the white portion cures on the vat. The maximum average intensity of the white portion is approximately 7.78 mW/cm$^2$. Hence each layer of 0.1 mm is cured under 6s, layer thickness less than 0.1 mm require a shorter exposure period. For manually placing the fibers

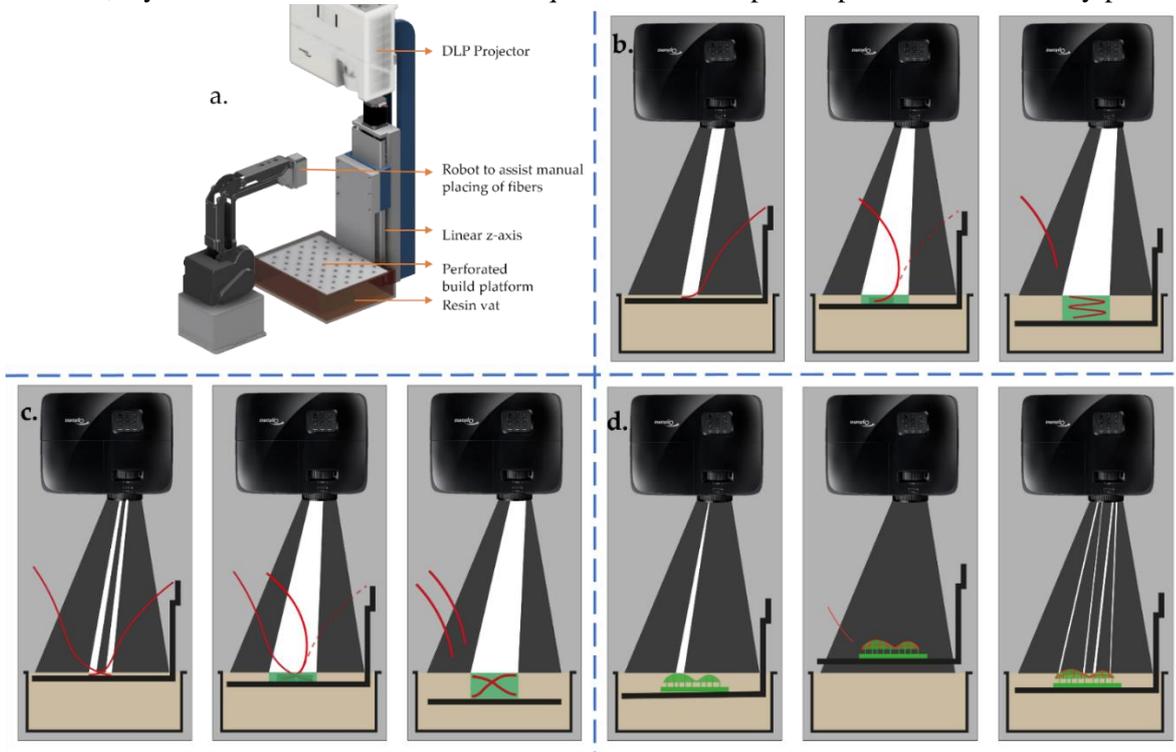

*Figure 2 a. Illustrates the 3D printing setup. b. Shows a three-step process of embedding a single helix track from left to right (red for fibers, green for green part), the left most shows a fiber-track being placed on the substrate, then slices are projected after the fibers are moved, right most shows end of the fiber track and subsequent layers being cured. c. In this three-step process two fibers are placed simultaneously instead of one in (b). d. Shows the first shell of the peanut being built, then without detaching the part, fibers are placed on the surface of this shell, right most step shows the projections of the second peanut shell cured on the fibers.*

during the solidification of a layer, we take help of a 4-DoF serial arm-robot to hold the fibers in tension after placement. The arm-robot is a Dobot MG400, four serial revolute configurations with the end of arm parallel to the X-Y work plane. For the initial placement of fiber, a point-laser of 405nm is used to rapidly cure the matrix around the fiber to the substrate. After the parts are built, they are thoroughly washed with iso-propyl alcohol followed by post-curing for 100 minutes at 45 °C. Figure 2a shows an illustration of the actual setup.

Next, we describe the printing process of parts with 3D fiber-tracks thoroughly embedded in the matrix domain.

## 3. Results and Discussion

In this section, we present the parts printed with our printer to show how single and multiple continuous fibers are embedded in the layer-by-layer process. We also show the effect of embedding fibers in specific patterns for improving mechanical stiffness, impregnation of matrix into the fiber tows, and discuss the capillary effect inside the fiber tows.

### 3.1. Single continuous fiber embedded in 3D

We printed a cylindrical part with a fiber embedded as a helical track inside the matrix domain. The print files to this part are realized as two STL files, one with the cylindrical domain with a cavity for the helical track and the second file is the helical coil with diameter greater than the diameter of fiber tow. The two files are sliced using a freeware slicer ChituBox$^{TM}$. A series of binary (black and white) images are obtained and are indexed in the z-axis to match the features. The series of images are projected using a python script that controls the projector and the build-platform z-axis motor. The slices from the second file are projected to fix the position of the fiber in a particular z-height before an entire layer is projected thus minimizing the shadows cast by the fibers. Figure 2b shows an illustration of the above. Figure 3a shows the part printed from the process.

### 3.2. Multiple continuous and discrete fiber tracks embedded in 3D

We printed a similar cylindrical sample, where instead of one single helical track, we embed two half-helical tracks of different centers. The slices of the part are constructed and orchestrated in similar fashion as described in Section 2.3. Figure 2c shows an illustration of the process of printing. We also printed a cylinder with three continuous and discrete fiber tracks that are unidentical to demonstrate the capability of simultaneous embedding of multiple fiber tracks. Figure 3b shows the 3D printed part from the process.

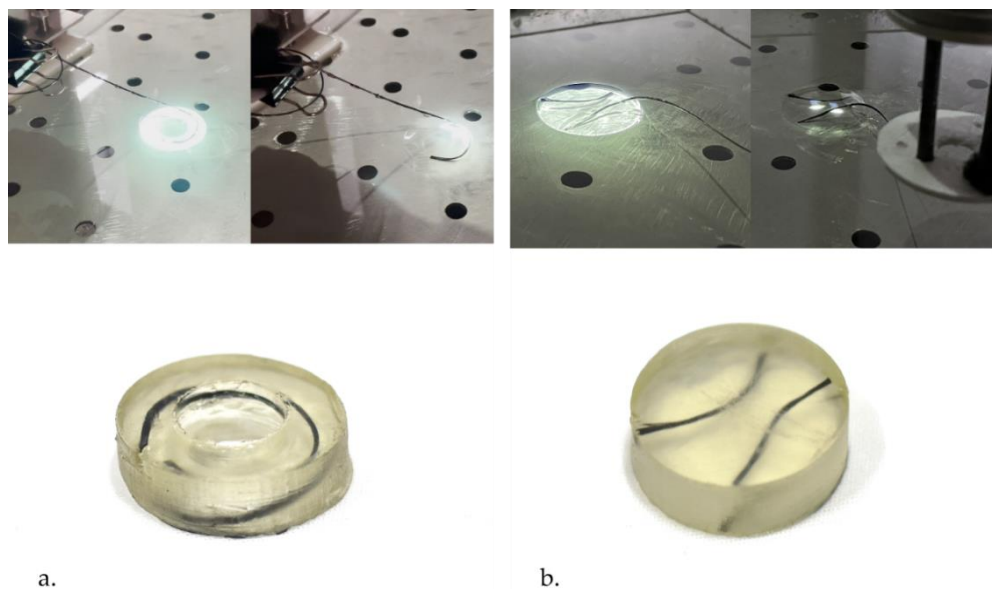

*Figure 3 the images from the top left to right of a and b shows the projections for entire layer and projection fixing fiber in their locations respectively, the bottom shows the part printed from this process.*

## 3.3. 3D printing parts with discrete continuous fibers inside a shell

Fibers on a peanut shell may appear as if they are present on the surface but the surface that contains fibers is embedded by matrix all around. Such structures are realized as two shells (part upon a part) separated by the surface that contains the fibers. First, the shell underneath the surface containing fibers is printed and then the fibers are placed on the printed part without detaching it from the build platform. After the fibers are securely placed, the upper shell is printed in similar fashion as the first. The projections are indexed with the built part for the second shell. Similarly, an example of a leaf is realized. Figure 2d shows an illustration of the process. Figure 4 shows a leaf and three peanut shells 3D printed with a shell-upon-shell based printing process, two with different fiber patterns and one without fiber. Here, the fiber looks like a tape or as if it is painted. This is because fiber bundle tends to splay during printing.

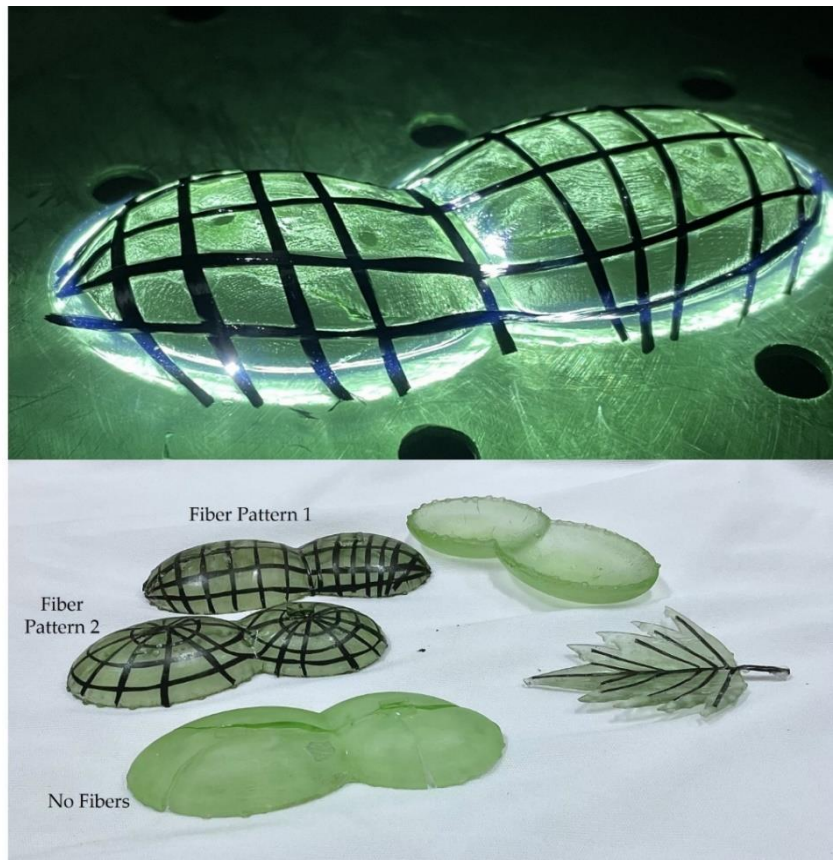

*Figure 4 Top images shows the cover shell being printed on the shell which has fibers placed on the top. Bottom images show peanut shaped parts with different fiber patterns and without it taken after compression test till failure (hence we put the broken parts together for this photo) and a leaf shaped part with fibers as a mid-rib is printed with the process explained in Section 3.3.*

## 3.4. Test for stiffness

3D printed carbon fiber reinforced structures are proven for strength in most of the literature, but mechanical stiffness is seldom explored. In the pursuit for a stiffer structure ideally, it is desired the structure takes large loads with negligible deflection. This idealistic structure is far from reality; a structure can be made stiffer for the loading conditions by optimal placement of fibers. We test the peanuts shaped parts printed from our printer, two with different fiber patterns and the third one without fibers for stiffness, for comparison.

We know that an actual peanut has a fiber arrangement like the fiber pattern 1 in Figure 4. We test the parts under compression between two plates. We find that the part without fibers gives highest stiffness than parts with fiber patterns. However, the part with fiber pattern 1: similar to an actual peanut is stronger than the pure polymer part. The fiber pattern 2 is an intuitive pattern of the principal stress to flowlines under the same loading conditions. We find the fibers arranged in the principal stress direction report lowest initial stiffness and highest failure strength than other fiber patterns. The part with fiber pattern 1 fails at a load of 1.4 kN whereas fiber pattern 2 does not fail under this loadcell capacity (2 kN), see Figure 5a. We also see that the fiber pattern 2 and part without any fibers show a similar trend of stiffness versus displacement. Fiber pattern 1 shows highest growth in stiffness unlike other parts, despite the fiber reinforced parts reported lower stiffness than pure polymer part, see Figure 5b. We note here that fibers contribute only to strength and not stiffness. On the contrary, fibers reduced the stiffness. This may be because of reduced stiffness of the fibers in the directions orthogonal to the axis of the fibers.

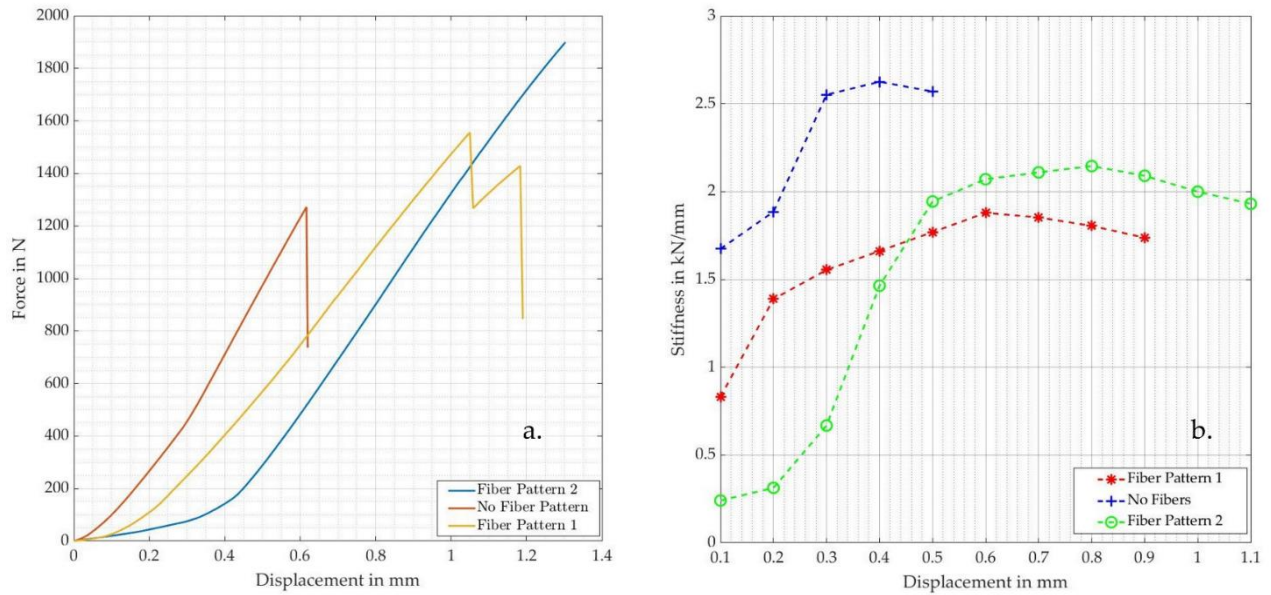

*Figure 5a. Force-Displacement plot of parts under compression. b. Stiffness derived from FD curve at 0.1 mm interval of deformation.*

### 3.5. Fiber adhesion and impregnation with matrix

Unlike FFF, fibers used in this study are tows of fibers, i.e., 3000 individual fibers form a fiber tow. Since these individual fibers are not adhered to one another, resin impregnation and subsequent polymerization adheres these fibers to the cured part. After flow of resin into the fiber bundle, the diameter of the fiber-track is nearly 10 times that of a layer height, hence any fiber track needs more than a two or three 0.1 mm-think layers to totally embed it. We place fiber tows in two ways to visualize the resin impregnation into the fiber tow. First, we place it parallel to build plane to see the resin impregnation laterally. Secondly, we place fibers in a helical track and then cut a small portion to see the impregnation in the fiber tow across the layers. We allow resin impregnation slowly and thoroughly as the part is built layer upon layer. We used a transmittance microscope to look at the interfaces. We find that fibers are embedded well inside the matrix in both cases. Figure 6 shows fibers embedded in a helical track; the red circle shows the fibers surrounded by cured resin. Figure 7 shows two views of fibers embedded parallel to the build plane.

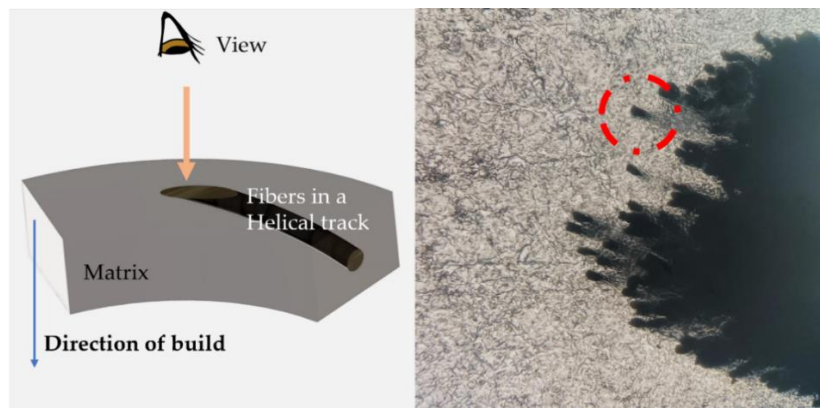

*Figure 6 Matrix impregnation along the helical fiber track. Each fiber is of 7μm diameter placed at an angle nearly 45° out-of-plane.*

Figure 7a shows the cross-section along the track and Figure 7b shows the lateral view of the fiber-track. We notice that there exist cavity marks of fibers that were displaced while sample preparation. This indicates that fibers were stuck there before it was displaced, thus providing evidence of surface contact, and wetting of fibers with liquid resin.

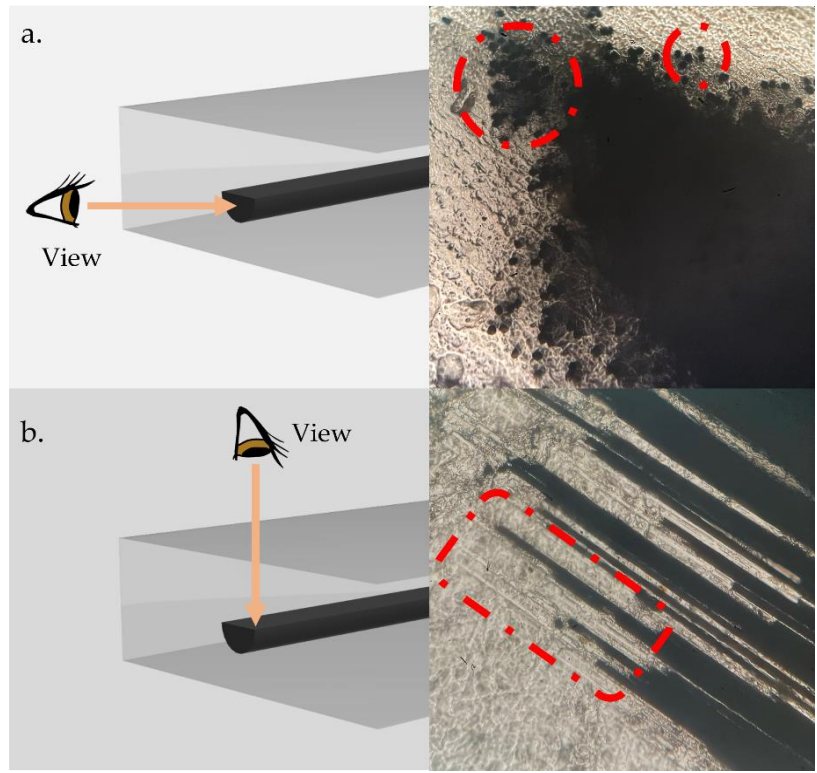

*Figure 7a. Shows the cross-sectional view of the fiber track b. shows the lateral view of the same. The red regions highlight the regions of interest.*

In the previous subsection, we checked the strength of the parts printed. The parts that failed do not display thorough impregnation of resin. This can be a virtue of the 3D printing process (Shell-upon-shell) and post processing. In post processing we wash the parts with Iso-Propyl Alcohol, which may remove uncured resin situated inside the carbon fiber bundle. The exposed carbon fiber tracks in the part forms the passage for IPA and removal of uncured resin. Figure 8 shows the fiber strands splayed near the crack in the peanut part with fiber pattern 1.

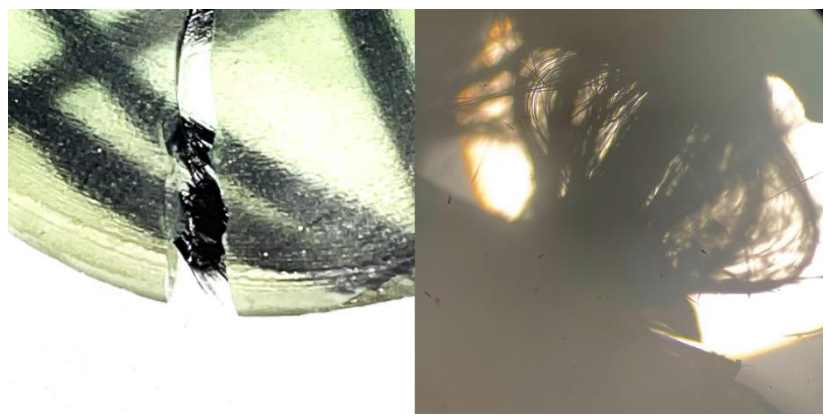

*Figure 8 Peanut with fiber pattern 1 failed under compression (left), the individual fiber strands seen under microscope, indicating lack of adhesion.*

## 4. Conclusions

State-of-the-art topology optimization techniques give designs for fiber-reinforced composites that have multiple continuous fibers positioned inside the matrix in specific spatial patterns. Such 3D designs cannot be realized in 3D printing yet even though it is possible to do it in parts with 2D fiber patterns. In this work, we presented techniques and a 3D printer that addressed this need. We demonstrated two examples of embedded fibers, one in a cylindrical tube with a single fiber in a helical pattern and the other in a solid cylinder with two fibers as separate helices. We also demonstrated fibers reinforced along a curved surface with the 3D printer developed as part of this work. We tested different fiber patterns embedded in a curved surface resembling half of a peanut shell under compressive load and found significant increase in strength as compared to the shell without fibers. We note that fiber-reinforcement is effective because fiber-matrix interface is strong, and the matrix is transferring loads effectively to the fibers. The fiber pattern on peanut shell justified that specific spatial fiber patterns indeed enhance strength. Topologically optimized designs with fiber-reinforcement show such fiber arrangements, and these structures can be realized with the 3D printer presented in this work.